\documentclass[amsmath,amssymb,nofootinbib,preprint]{revtex4}

\usepackage{latexsym,comment}
\usepackage{amssymb}
\usepackage{amsfonts}
\usepackage{amsmath,color}
\usepackage[dvips]{graphicx}
\usepackage{bm}

\def\ber{\begin{eqnarray}}
\def\eer{\end{eqnarray}}
\def\beq{\begin{equation}}
\def\eeq{\end{equation}}

\begin{document}

\title{Weak-Field Spherically Symmetric Solutions in $f(T)$ gravity}

\author{Matteo Luca Ruggiero}
\email{matteo.ruggiero@polito.it}
\affiliation{DISAT, Politecnico di Torino, Corso Duca degli Abruzzi 24,  Torino, Italy\\
 INFN, Sezione di Torino, Via Pietro Giuria 1, Torino, Italy}

\author{Ninfa Radicella}
\email{ninfa.radicella@sa.infn.it}
 \affiliation{Dipartimento di Fisica ÒE.R. CaianielloÓ, Universita` di Salerno, Via Giovanni Paolo II 132,  Fisciano (Sa), Italy\\
 INFN, Sezione di Napoli, Gruppo Collegato di Salerno,  Napoli, Italy}

\date{\today}

\begin{abstract}
We study weak-field solutions having spherical symmetry in $f(T)$ gravity; to this end, we solve the field equations for a non diagonal tetrad, starting from Lagrangian  in the form $f(T)=T+\alpha T^{n}$, where $\alpha$ is a small constant, parameterizing the departure of the theory from GR. We show that the classical spherically symmetric solutions of GR, i.e. the Schwarzschild and Schwarzschild-de Sitter solutions, are perturbed by terms in the form $\propto r^{2-2n}$ and  discuss the impact of these perturbations in observational tests.
\end{abstract}

\maketitle

\section{Introduction}\label{sec:intro}

Since the very discovery of the accelerated cosmic expansion \cite{Riess:1998cb,Perlmutter:1998np}, and its confirmation due to multiple observations \cite{Hinshaw:2012aka,Eisenstein:2005su,Wang:2008vja}, it has been customary to investigate theories  that extend general relativity (GR), in order to get an agreement with the observations, without requiring the existence of \textit{dark entities}. Hence, motivated by this instance, which recognizes that GR fails in describing gravity at large scales (consider also the old issue of the rotation curves of spiral galaxies \cite{Binney87}), several theories have been proposed to generalize  Einstein's theory. Some of these modified models of gravity are geometric extensions of GR, in other words they are based on a richer geometric structure, which is supposed to give the required ingredients to support the observations.

 As a prototype of this approach, one can consider the $f(R)$ theories,  where the gravitational Lagrangian depends on
a function $f$ of the curvature scalar $R$
(see \cite{Sotiriou:2008rp,Capozziello:2007ec} and references therein): when $f(R)=R$ the action reduces to the usual Einstein-Hilbert action, and Einstein's theory is obtained. Another example is given by the so-called $f(T)$ theories, which have similarities and differences with respect to $f(R)$. To begin with, they are based on teleparallel gravity (TEGR) \cite{pereira}, where the gravitational interaction is determined by torsion, and the torsion scalar $T$ appears in the Lagrangian instead of the curvature scalar.  Furthermore, the underlying Riemann-Cartan space-time is endowed with the Weitzenb\"ock connection (instead of the Levi-Civita connection), which is not commutative under the exchange of the lower indices, and has zero curvature while non-zero torsion. Actually, Einstein himself proposed such an alternative point of view on gravitation in terms of torsion and tetrads \cite{einstein25}. In fact, in the TEGR picture the tetrads field
is promoted to be the dynamical field instead of the metric tensor. In spite of these differences, TEGR and GR have equivalent dynamics: in other words, every solution of GR is also solution of TEGR.  However, when TEGR is generalized to $f(T)$  by considering a gravitational Lagrangian that is a function of the torsion scalar, the equivalence breaks down \cite{Ferraro:2008ey, Fiorini:2009ux}. As a consequence  $f(T)$ theories can be considered potential candidates for explaining (on a purely geometric ground) the accelerated expansion of the universe, without requiring the existence of exotic cosmic fluids (see e.g.\cite{cardone12}).

While $f(R)$ theories gives fourth order equations (at least in the metric formalism, while they are still second order in the Palatini approach, see e.g. \cite{Capozziello:2007ec}), the $f(T)$ field equations are second order in the field derivatives since the torsion scalar is a function of the square of the first derivatives of the tetrads field. Furthermore, as for $f(R)$ theories, the generalized TEGR displays additional degrees of freedom (whose physical nature is still under investigation \cite{li11}) related to the fact that the equations of motion are not invariant under local Lorentz transformations \cite{Li:2010cg}.  In particular, this implies the existence of a preferential global reference frame defined by the autoparallel curves of the manifold that solve the equations of motion. Consequently,  even though the symmetry can help in choosing suitable coordinates to write the metric in a simple way, this does not give any hint on the form of the tetrad.  As discussed in  \cite{tamanini12},  a diagonal tetrad -that could in principle be a good working-ansatz for dealing with diagonal metrics- is not a good choice to properly parallelize the spacetime  both in the context of  non-flat homogenous and isotropic cosmologies (Friedman-Lemaitre-Robertson-Walker universes) and in spherically symmetric space-times (Schwarzschild or Schwarzschild-de Sitter solutions).   
 
The case of the Schwarzschild solution and, more in general, of the spherically symmetric solutions in $f(T$) gravity, is particularly important, because these solutions, which describe the gravitational field of point-like sources, allow to test $f(T)$ theories at scales different from the cosmological ones, e.g. in the solar system.  {Such class of solutions -both with diagonal and non-diagonal tetrads- have been received much attention during the last few years, see for instance \cite{ss1,ss2,ss3,ss4,ss5,ss6,ss7}.}  Indeed, $f(T)$ theories can be used to explain the cosmic acceleration and  observations on large scales (e.g. via galaxy clustering and cosmic shear measurements \cite{Camera:2013bwa}), but we must remember that since GR is in excellent agreement  with  solar
system and binary pulsar observations \cite{Will:2005va},  every theory that aims at
explaining the large scale dynamics of the Universe  should reproduce GR  in a
suitable weak-field limit: the same holds true for $f(T)$ theories.  Recently, solar system data \cite{Iorio12, Xie:2013vua} have been used to constrain $f(T)$ theories; these results are based on the spherical symmetry solution found by Iorio\&Saridakis \cite{Iorio12}, who used a diagonal tetrad.  In this paper we follow the approach described in \cite{tamanini12} to define a ``good tetrad'' in $f(T)$ gravity - that is consistent with the equations of motion without constraining the functional form of the Lagrangian -  and solve the field equations to obtain weak-field solutions with a power-law ansatz for an additive term to the TEGR Lagrangian, $f(T)=T+\alpha T^{n}$.

This paper is organized as follows: in Section \ref{sec:fT-gravity} we review the theoretical framework of $f(T)$ gravity and write the field equations, whose solutions for spherically symmetric space-times, in weak-field approximation, are given in Section \ref{sec:wfsol}. Eventually, discussion and conclusions are in Sections \ref{sec:disc} and \ref{sec:conc}.

\section{$f(T)$ gravity field equations}\label{sec:fT-gravity}

We start by briefly discussing  the $f(T)$ gravity framework that leads to the field equations. To begin with, we point out that, in this scenario, the metric tensor can be viewed as a subsidiary field, and the vierbein field is the dynamical object whose components in a given coordinate basis $e^a_\mu$ are related to the metric tensor by
\beq
g_{\mu \nu}(x) = \eta_{a b} e^a_\mu(x) e^b_\nu(x) \ ,  \label{eq:gmunueta}
\eeq
where $\eta_{a b} = \text{diag}(1,-1,-1,-1)$. Notice that latin indexes refer to the tangent space while greek indexes label coordinates on the manifold. 
Hence, the dynamics is obtained by the action\footnote{We use units such as $c=1$.}

\begin{equation}
{\cal{S}} = \frac{1}{16 \pi G} \int{ f(T)\, e \, d^4x} + {\cal{S}}_M \ ,
\label{eq: action}
\end{equation}
where $e = \text{det} \  e^a_\mu = \sqrt{-\text{det}(g_{\mu \nu})}$ and ${\cal{S}}_M$ is the action for the matter fields\footnote{Notice that many authors write the gravitational Lagrangian  in the form $T+f(T)$, thus denoting the deviation from GR by means of the function $f(T)$: on the contrary, here $f(T)$ is the \textit{whole} Lagrangian.}. Here $f(T)$ is a differentiable function of the \textit{torsion scalar} $T$, which is defined as
\beq
T = S^\rho_{\ \mu \nu} T_\rho ^{\ \mu \nu} \ , \label{eq:deftorsions}
\eeq
where the \textit{contorsion tensor} $S^\rho_{\ \mu \nu}$ is defined by
\beq
S^\rho_{\ \mu \nu} = \frac{1}{4} \left ( T^{\rho}_{\ \ \mu \nu} - T_{\mu \nu}^{\ \ \rho}+T_{\nu \mu}^{\ \ \rho} \right ) +
\frac{1}{2} \delta^\rho_\mu T_{\sigma \nu}^{\ \ \sigma} - \frac{1}{2} \delta^\rho_\nu T_{\sigma \mu}^{\ \ \sigma} \ , \label{eq:defcontorsion}
\eeq
and the \textit{torsion tensor} $T^\lambda_{\ \mu \nu} $ is
\beq
T^\lambda_{\ \mu \nu} = e^\lambda_a \left( \partial_\nu e^a_\mu - \partial_\mu e^a_\nu \right ) \ . \label{eq:deftorsiont}
\eeq
Varying the action with respect to the vierbein $e^a_\mu(x)$, one gets the field equations

\beq
e^{-1}\partial_\mu(e\  e_a^{\ \rho}   S_{\rho}^{\ \mu\nu})f_T+e_{a}^{\ \lambda} S_{\rho}^{\ \nu\mu} T^{\rho}_{\ \mu\lambda} f_T
+  e_a^{\ \rho}  S_{\rho}^{\ \mu\nu}\partial_\mu (T) f_{TT}+\frac{1}{4}e_a^{\nu} f = 4\pi G e_a^{\ \mu} {\cal{T}}_\mu^\nu,
\label{eq: fieldeqs}
\eeq
where ${\cal{T}}^\nu_\mu$ is the matter energy\,-\,momentum tensor and subscripts $T$ denote differentiation with respect to $T$.

We look for spherically symmetric solutions of the field equations, so  we start from the metric
\begin{equation}
ds^2=e^{A(r)}dt^2-e^{B(r)}dr^2-r^2 d\Omega^2 \ , \label{metric}
\end{equation}
where $d\Omega^{2}= d\theta^2+ \sin^2 \theta d\phi^2$. Due to the lack of Local Lorentz Invariance, tetrads connected by local Lorentz transformations lead to the same metric - i.e. the same causal structure - but different equations of motions, thus physically inequivalent solutions. This means that, even in the spherically symmetric case, for which symmetry helps us in choosing the coordinates and the metric tensor in a simple form, it is quite complicated to do an ansatz for the tetrad field. In particular, for the symmetry and coordinates we are dealing with, it turns out to be a mistake to choose a diagonal form for $e^a_{\ \mu}$: it does not properly parallelize the static spherically symmetric geometry in the context of $f(T)$ gravity. 

Then, with this {\sl caveat} in mind, it is possible to derive the field equations for the non diagonal tetrad:

$$
e_\mu^a=\left( \begin{array}{cccc}
e^{A/2}         &   0                                             &   0                                            &    0         \\
0                 &e^{B/2} \sin{\theta}\cos{\phi}   & e^{B/2} \sin{\theta}\sin{\phi}&  e^{B/2} \cos{\theta}\\
0                 &-r \cos{\theta}\cos{\phi}   & -r  \cos{\theta}\sin{\phi}&  r \sin{\theta}\\
0                 &r \sin{\theta}\sin{\phi}   & -r  \sin{\theta}\cos{\phi}&  0\\
 \end{array} \right) 
$$
following the approach described in \cite{tamanini12}: in doing so,  the functional form of the Lagrangian and the specific form of the torsion scalar are not  constrained {\sl a priori}.  {We remark once more that different choices of tetras, while giving back the same metric represent different physical theories. In this work we are interested in a specific tetras that does not lead to a constant torsion scalar. In such a theory  Birkhoff theorem does hold, as shown in \cite{tamanini12} while the most general vacuum solution is not of Scwharzschild De Sitter kind, as it happens with tetrads for which $T=\text{const}$. Then, it is worthwhile to investigate the features of the spherically symmetric solutions in this case for a generic Lagrangian, that nevertheless should admit the Schwarzschild solution when it reduces to the teleparallel equivalent of GR.} The field equations are
\begin{eqnarray}
&&\frac{f(T)}{4}-f_T\frac{e^{-B (r)}}{4r^2}\left(2-2e^{B(r)}+r^2 e^{B(r)} T -2r B'(r)\right)+\nonumber\\
&&-f_{TT} \frac{T'(r) e^{-B(r)}}{r}\left(1+e^{B(r)/2}\right)=4\pi \rho \label{00eq1}\\
&&-\frac{f(T)}{4}+f_T\frac{e^{-B (r)}}{4r^2}\left(2-2e^{B(r)}+r^2 e^{B(r)} T -2r A'(r)\right)=4 \pi p \label{11eq1}\\
&&f_T\left[-4+4e^{B(r)}-2r A'(r)-2r B'(r)+r^2A'(r)^2-r^2A'(r)B'(r)+2r^{2}A''(r)\right]+\nonumber\\
&&+2rf_{TT}T'\left(2+2e^{B(r)/2}+rA'(r)\right)=0 \label{eq31}
\end{eqnarray}
where $\rho,\ p$ are the energy density and pressure of the matter energy-momentum tensor, and the prime denotes differentiation with respect to the radial coordinate $r$. Moreover, the torsion scalar is
\begin{equation}\label{torsionscalar}
T=\frac{2e^{-B(r)}(1+e^{B(r)/2})}{r^{2}}\left[1+e^{B(r)/2}+r A'(r)\right].
\end{equation}


\section{weak-field solutions} \label{sec:wfsol}

Exact solutions in vacuum ($\rho=p=0$) and in presence of a cosmological constant ($\rho=-p$) of the above field equations are thoroughly discussed in \cite{tamanini12}; here we are interested in weak-field solutions with non constant torsion scalar, i.e. $T'=dT/dr \neq 0$. 

 Indeed,  for actual physical situations such as  in the solar system, the gravitational field is expected to be just a small perturbation of a flat background Minkowski spacetime. As a consequence, we write 
\beq
e^{A(r)}=1+A(r), \quad e^{B(r)}=1+B(r) \label{eq:devAB}
\eeq
and confine ourselves to linear perturbations.  Moreover, we consider Lagrangians  of sufficient generality,  that we write in the form $f(T)=T+\alpha T^n$, where $\alpha$ is a small constant, parameterizing the departure of these theories from $GR$, and $|n| \neq 1$.

To begin with, we consider the case $n=2$, that has been already analyzed in \cite{Iorio12}. From (\ref{00eq1}-\ref{eq31}) we obtain the following solutions
\beq
A(r)=-32\,{\frac {\alpha}{{r}^{2}}}-{\frac {{\it C_{1}}}{r}} \label{eq:sola}
\eeq
\beq
B(r) =96\,{\frac {\alpha}{{r}^{2}}}+{\frac {{\it C_{1}}}{
r}} \label{eq:solb}
\eeq
where $C_{1}$ is an integration constant. Then, on setting $C_{1}=2M$, we get the weak-field limit of the Schwarzschild solution plus a correction due to $\alpha$:
\beq
ds^2=\left(1-\frac{2M}{r} -32 \frac{\alpha}{r^{2}}\right)dt^2-\left(1+ \frac{2M}{r} +96 \frac{\alpha}{r^{2}} \right)dr^2-r^2 d\Omega^2 \label{eq:metricasol1}
\eeq

Eventually, the torsion scalar turns out to be
\beq
T(r)=\frac{8}{r^{2}}-128\frac{\alpha}{r^{4}}. \label{eq:torsionscalarsol}
\eeq

These results can be compared to those obtained in \cite{Iorio12}, where a Lagrangian in the form $f(T)=T+\alpha T^{2}$ was considered.  While to lowest order approximation in both cases the perturbations are proportional to $1/r^{2}$, the numerical coefficients are different: this is not surprising, since the authors in \cite{Iorio12} solve different field equations.  In particular they use a diagonal tetrad, 
which constrains the torsion scalar to be constant (see e.g. \cite{tamanini12} and references therein): however, the solution given in \cite{Iorio12} does not seem to have a constant torsion scalar, which makes it  inconsistent.

Likewise, if we look for solutions of the equations (\ref{00eq1})-(\ref{eq31}) with $\rho=k$, $p=-k$, which corresponds to a cosmological constant, we obtain
\beq
ds^2=\left(1-\frac{2M}{r} -32 \frac{\alpha}{r^{2}}-\frac 1 3 \Lambda r^{2}\right)dt^2-\left(1+ \frac{2M}{r} +96 \frac{\alpha}{r^{2}}+\frac 1 3 \Lambda r^{2} \right)dr^2-r^2 d\Omega^2 \label{eq:metricasol1k}
\eeq
where we set $k=\frac{\Lambda}{8\pi}$ and $\Lambda$ is the cosmological constant. The torsion scalar is the same as (\ref{eq:torsionscalarsol}). So, the weak-field limit of  Schwarzschild - de Sitter solution is perturbed by terms that are proportional to $\alpha$.

The previous results can be generalized to the case of a Lagrangian in the form $f(T)=T+\alpha T^n$, and we get
\beq
A(r)=-\frac{C_{1}}{r}-\alpha \frac{r^{2-2n}}{2n-3}2^{3n-1} -\frac 1 3 \Lambda r^{2}
\label{eq:solakn}
\eeq

\beq
B(r)=\frac{C_{1}}{r}+\alpha \frac{r^{2-2n}}{2n-3} 2^{3n-1} \left(-3n+1+2n^{2} \right)+\frac 1 3 \Lambda r^{2}
\label{eq:solbkn}
\eeq
In particular, if $\Lambda=0$, we obtain vacuum solutions.  Notice that, on setting $C_{1}=2M$, we obtain a weak-field Schwarzschild - de Sitter solution  perturbed by terms that are proportional to $\alpha$ and decay with a power of the radial coordinate, the specific value depending on the power-law choosed in the Lagrangian. The torsion scalar is
\beq
T(r)=\frac{8}{r^{2}}+2\alpha r^{-2n}2^{3n} \left(n+1 \right) \label{eq:torsionnk}
\eeq
while the perturbation terms due to the deviation from GR are in the form
\beq
A_{\alpha}(r)=\ \alpha a_{n} r^{2-2n}, \quad B_{\alpha}(r)= \alpha b_{n} r^{2-2n}  \label{eq:defABa}
\eeq
where $\displaystyle a_{n}= \frac{2^{3n-1}}{2n-3}, \, b_{n}=\frac{2^{3n-1}}{2n-3} \left(2n^{2}-3n+1 \right)$. A close inspection of the perturbation terms reveals that they go to zero both when $r \rightarrow \infty$ with $n > 1$ and  when $r \rightarrow 0$ with $n<1$. In the latter case,  in order the keep the perturbative approach self-consistent, a maximum value of $r$ must be defined  to consider these terms as perturbations of the flat space-time background.  

We remark here that our linearized approach can be applied to arbitrary polynomial corrections to the torsion scalar: as a consequence, by writing an arbitrary function as a suitable power series, it is possible to evaluate its impact  as a perturbation of the weak-field spherically symmetric solution in GR: the n-th term of the series gives a contribution proportional to $r^{2-2n}$.

 {It could be interesting to test the impact of the perturbations (\ref{eq:defABa}). To this end, we remember that it is possible to obtain the secular variations of the Keplerian orbital elements due to general spherically symmetric perturbations of the GR solution, describing the gravitational field around a point-like mass, as one of us showed in \cite{Ruggiero:2010yn}. For instance, the average over one orbital period of the secular precession of pericenter turns out to be:
\beq
<\dot{\omega}>=\frac 1 4 \alpha \frac{2^{3n-1} \left(2n-2 \right)\left(1-e^{2}\right)^{3-2n}}{n_{b}a^{2n}}\, \mathrm{F}\left(2-n,\frac 5 2 -n, 2,e^{2} \right), \quad \mathrm{for} \  n>\frac 3 2 \label{eq:omegadot1}
\eeq
\beq
<\dot{\omega}>=\frac 1 4 \alpha \frac{2^{3n-1}\left(2-2n \right)\left(3-2n \right)\sqrt{1-e^{2}}}{\left(2n-3 \right)n_{b}a^{2n}}\, \mathrm{F}\left(n, n- \frac 1 2 , 2, e^{2} \right), \quad \mathrm{for} \  n \leq \frac 1 2 \label{eq:omegadot2}
\eeq
In the above equations $n_{b}$, $a$, $e$ are, respectively, the mean motion, the semi major axis and the eccentricity of the unperturbed orbit, while $\mathrm F$ is the hypergeometric function. These relations can be used to constrain the parameters $\alpha,n$, on the bases of the ephemerides data.}

\section{Discussion} \label{sec:disc}

It is useful to comment on  the constraints one can infer for the parameters of our model from solar system data. But, before proceeding, it is important to emphasize a point about the tests of $f(T)$ gravity. In theories with torsion, there is a sharp distinction between the test particles trajectories: \textit{autoparallels,} or affine geodesics, are curves along which the velocity vector is transported parallel to itself, by the space-time connection; \textit{extremals,} or metric geodesics, are  curves of extremal space-time interval with respect to the space-time metric \cite{Mao:2006bb}. While in GR autoparallels and extremals curves do coincide and we can simply speak of geodesics,  the same is not true when torsion is present.  So, it is not trivial to define the actual trajectories of test particles. The results obtained by \cite{Iorio12} and \cite{Xie:2013vua},  {together with the expressions (\ref{eq:omegadot1}) and (\ref{eq:omegadot2}) of the secular precession of the pericenter,} strictly apply to the case of metric geodesics. According to us, this is a very important issue, that is often neglected in the literature pertaining to theories alternative to GR based to torsion: we will focus on this issue in a forthcoming publication \cite{RCCR}.  {In the same publication, we will constrain the parameters $\alpha$ and $n$, taking into account the recent data of the ephemerides of the Solar System provided by INPOP10a \cite{fienga10,fienga11} and EPM2011 \cite{pit1,pit2,pit3}. Actually, perturbations  in the form of power-law are present in different models of modified gravity, and their impact on the Solar System dynamics has been analyzed, for instance, in \cite{Iorio:2007ee, Adkins:2007et, liu}.}

Bearing this in mind,  it is possible to comment on our results and compare them to those already available in the literature  {pertaining to $f(T)$ theories.} In particular, because of the different choice of the tetrad, our solution, even in the case of a quadratic deformation of the TEGR Lagrangian, differs from the one found by Iorio\&Saridakis.   Both corrections are proportional to $1/r^{2}$,  but they have different  numerical coefficients. 

 {In particular, on substituting $n=2$ in Eq. (\ref{eq:omegadot1}), we obtain
\beq
<\dot{\omega}>= \frac{16 \alpha}{a^{4}n_{b}\left(1-e^{2}\right)} \label{eq:omegadot22}
\eeq
On the contrary, the corresponding expression obtained by Iorio\&Saridakis \cite{Iorio12} is
\beq
<\dot{\omega}>_{\mathrm{IS}}= \frac{3 \alpha}{a^{4}n_{b}\left(1-e^{2}\right)} \label{eq:omegadotIS}
\eeq
We see that they differ for a factor 16/3: the same happens to the constraints that can be obtained from our solution, by applying the approach described in \cite{Iorio12} and \cite{Xie:2013vua}.}

In particular, Iorio\&Saridakis \cite{Iorio12} derive constraints from the rate of change of perihelia of the first four inner planets, obtaining
\begin{eqnarray*}
|\Lambda|&\leq& 6.1\times 10^{-42} \ m^{-2}\\
|\alpha|&\leq& 1.8 \times 10^4 \ m^2.
\end{eqnarray*}
Tighter results have been obtained in a subsequent paper \cite{Xie:2013vua}, where the authors consider upper bounds deriving from different phenomena: perihelion advance, light bending, gravitational time delay  {\cite{sis1,sis2,sis3,sis4}}. But the strongest constraints come from the perihelion advance, in particular from some supplementary advances constructed by considering that the effects due to the Sun's quadrupole mass moment might represent possible unexplained parts of perihelion advance in GR \cite{fienga10}. This gives
\begin{eqnarray*}
|\Lambda|&\leq& 1.8	\times 10^{-43} \ m^{-2}\\
|\alpha|&\leq& 1.2 \times 10^2 \ m^2.
\end{eqnarray*}
The upper bound for the solution in Eqs. (\ref{eq:metricasol1k}) would be  {3/16 smaller, that is  $|\alpha| \leq 2.3 \times 10  \ m^{2}$.}

 {Eventually, we comment on the issue of the parameterized post-Newtonian formalism (PPN), in the framework of $f(T)$ gravity.  In order to test theories of gravity that give rise to  detectable torsion effects in the Solar System, a theory-independent formalism that generalizes the PPN formalism when torsion is present was developed in \cite{Mao:2006bb} (see also \cite{March:2011ry}): starting from symmetry arguments, the metric and the connection around a massive body are perturbatively  expressed in terms of dimensionless parameters related to the matter-energy content of the source, namely its mass and its angular momentum per unit mass. In doing so, the new parameters, which add to the original PPN ones, can be constrained by the experiments. Our results, however, cannot be directly described in this framework: an inspection of our solutions (\ref{eq:defABa}) clearly shows that the perturbations  \textit{are not related to the matter-energy content of the source,} rather they depend on $\alpha$, which  parameterizes the departure of the $f(T)$ theory from GR (see e.g. Eq. (4.2) in \cite{March:2011ry}). So, a new formalism is required to test the content of the Lagrangian by means of observations: in a sense, $\alpha$ can be considered a \textit{new} post-Newtonian parameter of this formalism. }

\section{Conclusions}\label{sec:conc}

We studied spherically symmetric solutions in the weak-approximation of $f(T)$ gravity. In particular, we started from a Lagrangian in the form $f(T)=T+\alpha T^{n}$, with $|n| \neq 1$,  where $\alpha$ is a small constant which parameterizes the departure of these theories from GR, and solved the field equations using a non diagonal tetrad, showing that, to lowest approximation order, the perturbations of the corresponding GR solutions (Schwarzchild or Schwarzchild- de Sitter) are in the form $\propto \alpha\  r^{2-2n}$. These results can be used to  evaluate the impact of the non linearity of the Lagrangian, for instance in the solar system.

The case $n=2$, corresponding to the Lagrangian $f(T)=T+\alpha T^{2}$ has been already analyzed by \cite{Iorio12} and \cite{Xie:2013vua}, who used solar system observations to set constraints on the parameter $\alpha$.  It is important to point out that the latter results are based on the solution obtained by Iorio\&Saridakis \cite{Iorio12}, where a diagonal tetrad was used, which forces the torsion scalar to be constant: however, that solution does not seem to have a constant torsion scalar, which makes the consequent constraints not reliable. 

On the other hand, since because of the invariance properties of $f(T)$ the choice of the tetrad field is crucial, we performed our calculations by using a more general non-diagonal tetrad, according to the prescriptions given in \cite{tamanini12}, and obtained  a new solution of the $f(T)$ quadratic model, for which the torsion scalar is not forced to be zero.

We used the results already available in the literature to obtain the correct constraints from solar system data on the $\alpha$ parameter for the Lagrangian $f(T)=T+\alpha T^{2}$,  even if we  pointed out that the distinction between autoparallels, or affine geodesics, and extremals, or metric geodesics, is crucial in $f(T)$ gravity, and deserves further investigation that we are going to carry out in forthcoming publications.

\acknowledgments

 The authors acknowledge support by Istituto Nazionale di Fisica Nucleare (INFN) and by the Italian Ministero dell'Istruzione dell'Universit\`a e della Ricerca (MIUR).

\end{document}